# Deep-UV-enhanced supercontinuum generated in tapered gas-filled photonic crystal fiber

**MALLIKA IRENE SURESH[1,2*], JONAS HAMMER[1], NICOLAS Y. JOLY[2,1], PHILIP ST.J. RUSSELL[1], AND FRANCESCO TANI[1,*]**

[1]Max Planck Institute for the Science of Light and  [2]Department of Physics, Friedrich-Alexander-Universität, Staudtstr. 2, 91058 Erlangen, Germany
*Corresponding authors: mallika-irene.suresh@mpl.mpg.de, francesco.tani@mpl.mpg.de

**We present the use of linearly down-tapered gas-filled hollow-core photonic crystal fiber in a single-stage, pumped with pulses from a compact infrared laser source, to generate a supercontinuum carrying significant spectral power in the deep ultraviolet (200 – 300 nm). The generated supercontinuum extends from the near infrared down to ~213 nm with up to 0.83 mW/nm in the deep ultraviolet.**

## 1. Introduction

Supercontinuum (SC) sources that extend from the ultraviolet (UV) to the infrared (IR) have numerous applications, for example in astronomy, microscopy, spectroscopy, optical coherence tomography and biological imaging [1–4]. Optical fibers provide an ideal platform for generating such broadband light, offering long interaction lengths and delivering the SC directly from the fiber with a well-defined spatial profile [5]. However, in order to efficiently generate light extending to the UV, standard silica fibers are not the platform of choice because of photodarkening effects in fused silica, which severely limits the lifetime of operation. Fluorozirconate glass photonic crystal fibers (PCFs) have been used to generate a SC extending down to a record 200 nm, but difficulties in obtaining suitable glass have brought a halt to this research [6]. Single-ring (SR) hollow-core PCFs filled with gas, which guide by anti-resonant reflection, are highly suitable candidates due to broadband guidance, low loss and low overlap between the light guided in the core and the thin silica walls in the cladding [7]. The dispersion and nonlinearity in these PCFs can be readily controlled by choosing the species and pressure of the gas filling the core. Tapering the fiber along its length offers an additional degree of freedom that turns out to be advantageous for extending the SC out to higher frequencies [8] and to lower the pump pulse energy [9]. Previously this was explored in solid-core PCFs to shift the zero dispersion wavelength to shorter wavelengths and allow a broader SC spectrum to be generated [8]. Tapering has also been numerically studied as a route to generating light in the extreme UV using gas-filled fibers [10].

Here we report the use of a down-tapered hollow core SR-PCF to generate a SC with significant spectral power in the deep UV (DUV). Tapering these PCFs for nonlinear optics has been recently explored for fabricating small-mode-area SR-PCFs [11] and used for low-energy-threshold generation of DUV light [12]. As shown by the numerical studies in [10, 13] as well as in our own simulations, an appropriately-designed transition from the untapered to the tapered section improves the efficiency of conversion to the UV.

In the experiments reported here, IR pulses carrying 4.7 W average power are coupled into the untapered end of the PCF, and undergo nonlinear spectral broadening within the fiber. The SC generated in this manner extends well into the DUV (~213 nm) with spectral power densities between 0.5 mW/nm and 0.83 mW/nm.

## 2. Taper design and fabrication

When designing an appropriate taper profile for SC generation, the most important parameter is the taper angle, which needs to be chosen not only to satisfy the adiabaticity criterion [14] to minimize coupling of pump light into higher order modes, but also to be gentle enough to allow efficient transfer of power to the UV, as discussed in [13]. The taper angle used in the simulations and experiments discussed here is constant at ~0.3 mrad along the entire transition, i.e., the taper profile is linear. This is around forty times smaller than the highest local taper angle at which the adiabaticity condition is satisfied.

Fig. 1 shows numerical simulations of IR pulses centered at 1030 nm, full-width half-maximum (FWHM) pulse duration of 211 fs, carrying 13 µJ of energy propagating through (a) an untapered fiber (core diameter 58 µm, 8 cladding capillaries of diameter 23 µm and thickness 330 nm) and (b) the proposed tapered fiber which is tapered to 46% of its original dimensions (taper ratio $T_R$=0.46), each filled with 15 bar of argon. The taper profile can be gleaned from the shift in the zero-dispersion wavelength, which follows the taper dimensions; in the untapered case it is at 914 nm across the entire length while in the tapered fiber it shifts down to 626 nm over a transition length of 6 cm, starting at the 60 cm point. As can be seen, the SC extends down to 380 nm in the untapered fiber, while in the tapered fiber, the shifting dispersion landscape allows generation of multiple dispersive waves at shorter wavelengths, resulting in additional spectral broadening down to 224 nm. The simulations include the effect of anti-crossings between the core mode and resonances in the glass walls, which cause high narrow-band loss peaks and alter the dispersion. Tapering causes the capillary walls to become thinner, which shifts these loss peaks to shorter wavelengths [11], as shown in Fig. 3(a), where the loss spectra are plotted for different taper ratios using the empirical expression in [15]. We do not consider thermal effects, which can be important at high average powers [16].

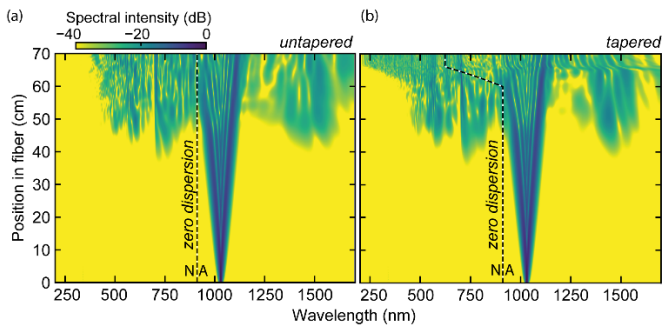

Fig. 1. Numerical simulations showing the spectral evolution of pulses (FWHM duration 211 fs, energy 13 µJ) initially centered at 1030 nm, propagating in 15 bar of argon in (a) untapered SR-PCF and (b) the same fiber tapered to 46% of its original dimensions over 6 cm from the 60 cm point onwards. The dashed lines mark the zero dispersion wavelength separating N: normal and A: anomalous dispersion regions.

The tapers were fabricated using the standard sweeping flame technique in which an oxy-butane flame gently brushes across a length of fiber held between two computer-controlled stages, that are programmed to move apart as the fiber softens when heated. As described in [11], during tapering the capillaries were kept at atmospheric pressure and the core at 70 mbar, thus counter-balancing surface-tension-driven collapse of the molten capillaries. The fabricated taper was cleaved at the waist to remove the up-tapered transition region and then used in the experiments discussed below.

### 3. Supercontinuum generation

Fig. 2(a) shows the experimental set-up in which the fabricated taper is used for single-stage SC generation. A commercial compact laser source (Light Conversion Carbide system: central wavelength at 1030 nm, FWHM duration of 211 fs and repetition rate tunable from 60 kHz to 1 MHz) was used to pump the tapered SR-PCF, which was mounted in a gas cell filled with 15 bar of argon. The SC spectrum measured at the fiber output was optimized by slightly adjusting the duration and chirp of the laser pulses. The pulse energy was controlled using a combination of a half-waveplate and a thin film polarizer. The two insets show scanning electron micrographs of the PCF structure in the untapered region and at taper waist, showing isomorphic downscaling to 46% of the original size.

The pulses carrying 13 µJ (soliton order 62 at the input) undergo modulational instability as they propagate through the fiber and generate a SC that reaches 23 nm (at −25 dB from the peak) in 15 bar of argon, as shown by the purple curve in Fig. 2(b); the unshaded area corresponds to the part of the spectrum measured by an optical spectrometer (Avantes 2048XL) and the shaded area to that measured by an optical spectrum analyzer (Yokogawa AQ16315A). The SC bandwidth is smaller than predicted in the simulations, which we attribute to differences in the optical loss and dispersion due to the non-ideal geometry of the SR-PCF, caused by slight variations in wall thickness between capillaries and a slightly elliptical core. Using the full power of the laser yielded power of 4.7 W in the fiber (at a repetition rate of 350 kHz this gives a pulse energy of 13 µJ). The power in the DUV was estimated by first integrating the spectrum from 200 to 300 nm and then calibrating it to the total output power, measured using a thermal power meter after the uncoated $MgF_2$ window at the output of the gas cell. In the DUV between 200 and 300 nm, the spectral power density at 350 kHz was found to be 0.69 mW/nm, with 95% of the output power linearly polarized along one axis. Adjusting the pulse energy and gas pressure to generate even higher frequencies, we obtained a SC extending down to 214 nm (blue curve in Fig. 2(b)) in 11 bar of argon at a repetition rate of 288 kHz (same average power) and a spectral power density of 0.5 mW/nm in the DUV.

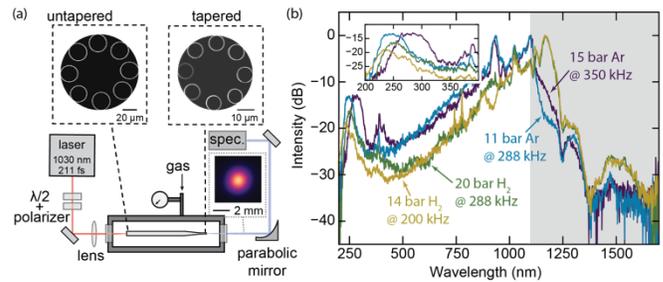

Fig. 2. (a) Experimental set-up: Pulses (4.7 W, 1030 nm, FWHM 211 fs) are launched into the tapered SR-PCF and the generated spectrum is collimated with a parabolic mirror before being measured. Above are scanning electron micrographs of the PCF structure at the untapered and tapered ends, and the color plot shows the intensity profile of the SC captured on a visible-near-IR beam profiler. (b) SC obtained at different gas pressures and repetition rates (shaded area measured by an optical spectrum analyzer, unshaded by an optical spectrometer). Inset zooms into the UV (200-400 nm).

These results are further improved if argon is replaced with hydrogen, as shown by the green and yellow curves in Fig. 2(b). The dynamics for 16 µJ pulses and 20 bar of $H_2$ are similar to those for 13 µJ pulses and 15 bar of argon. As reported in [17,18], optically excited Raman coherence phase-modulates the DUV dispersive waves, resulting in generation of wavelengths shorter than achievable using noble gases. As a result the SC extends down to 220 nm in 20 bar of $H_2$ with a power density of 0.83 mW/nm at 288 kHz in the DUV (green curve). An additional advantage of using $H_2$ is its higher thermal conductivity [16], which makes it possible to operate at higher average powers. On the other hand, only 80% of the output power was along one polarization axis, because of alterations in polarization state caused by rotational Raman scattering. After adjusting the pulse energy and gas pressure to reach still higher frequencies, we obtained a SC extending down to 213 nm with spectral power density of 0.58 mW/nm in the UV (repetition rate 200 kHz, $H_2$ pressure 14 bar, yellow curve). The far-field of the output mode was imaged on a visible-near-IR CMOS camera (WinCamD-LCM), showing an excellent Gaussian-like beam profile and suggesting that the near-field was in the $LP_{01}$-like mode of the fiber (see inset Fig. 2 (a)).

A SC source is most useful if it is long-term stable. Over a 30 minute period (Fig. 3(b)), we found that the spectral bandwidth reduced and the total power increased by 1.4%, which we attribute to thermal effects in the set-up. We could however recover the full bandwidth by making small adjustments to the launching stages.

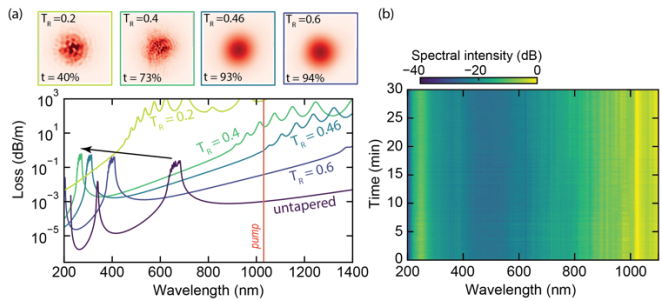

Fig. 3. (a) Upper panel: experimentally-observed intensity profiles of the modes at the pump wavelength (low power) in different tapers. Lower panel: empirically-calculated loss spectra [15] for different taper ratios, showing how the anti-crossing loss peaks shift to shorter wavelength. Note the high loss at the pump wavelength (vertical red line) for small taper ratios. (b) Stability of the generated SC over half an hour.

## 4. Discussion and conclusions

The use of linearly down-tapered gas-filled SR-PCF provides a simple means of enhancing the spectral power in the DUV (up to 0.83 mW/nm), while extending the spectrum to shorter wavelengths (down to 213 nm). A DUV spectral power of 50 mW between 200 and 300 nm could be obtained for a total SC power 2.5 W. Compared to previous work on untapered hollow-core fibers [19,20] the spectra reported here extend to significantly higher frequencies and carry several times higher power.

Numerical simulations predict that the SC can be extended even further into the UV for smaller taper waists. Thinning down of the structural features also results, however, in an increase in loss, especially at longer wavelengths. Indeed, at the smallest taper ratio, the transmission of the pump light was very poor, and the modal intensity profile distorted (upper panel of Fig. 3(a)). The spectral width of the first transmission window also depends on variations in the diameter of the capillaries surrounding the core and their wall thickness. Therefore, optimizing the fiber structure design and further improving the uniformity of the capillaries would lower the loss at long wavelengths and lead to a broader SC carrying even more power.

In order to generate broad and flat spectra in the DUV, it is necessary to operate in the MI regime at high soliton order. Given the relatively long duration of the pump pulses, it was necessary to tune the experimental parameters so as to access this regime, with the result that the generated SC exhibit shot-to-shot fluctuations. Such fluctuating spectra are interesting for extending noise-aided imaging and spectroscopic techniques into the DUV [21–23]. Scaling the system to higher powers and repetition rates with a suitable laser source would make it suitable for DUV photolithography, provided detrimental thermal effects are reduced by adding helium to the gas, to increase the thermal conductivity and the heat dissipation rate [24].

**Disclosures.** The authors declare no conflicts of interest.

**Funding.** M. I. Suresh acknowledges the Erlangen Graduate School in Advanced Optical Technologies (SAOT). Content in the funding section will be generated entirely from details submitted.